\documentclass[11pt, a4paper]{elsarticle}

\usepackage[UKenglish]{babel}
\usepackage{etoolbox}
\usepackage{mathtools}
\usepackage{epstopdf}
\usepackage{float}

\usepackage[nodayofweek]{datetime}

\usepackage[left=1.0in, right=1.0in, top=0.86in, bottom=0.86in]{geometry}

\newcommand{\figurewidth}{0.9\textwidth} 

\makeatletter
 \def\ps@pprintTitle{
 \let\@oddhead\@empty
 \let\@evenhead\@empty
 \def\@oddfoot{\centerline{\thepage}}
 \let\@evenfoot\@oddfoot}
\makeatother

\patchcmd{\emailauthor}{(#2)}{}{}{}

\patchcmd{\MaketitleBox}{\footnotesize\itshape\elsaddress\par\vskip36pt}{}{}{}
\patchcmd{\pprintMaketitle}{\footnotesize\itshape\elsaddress\par\vskip36pt}
{\footnotesize\itshape\elsaddress\par\parbox[b][36pt]{\linewidth}
{\vfill\hfill\textnormal{DESY Report 16-168 (Internally reviewed; approved for publication \today)}\hfill\null\vfill}}{}{}

\begin{document}

\begin{frontmatter}

\title{\textbf{Empirical Optimization of Undulator Tapering at FLASH2 \\ and Comparison with Numerical Simulations}}

\author[MAXaddress]{Alan Mak\corref{mycorrespondingauthor}}
\cortext[mycorrespondingauthor]{Corresponding author}
\ead{alan.mak@maxiv.lu.se}

\author[MAXaddress]{Francesca Curbis}
\author[DESYaddress]{Bart Faatz}
\author[MAXaddress]{Sverker Werin}

\address[MAXaddress]{MAX IV Laboratory, Lund University, Fotongatan 2, S-22594 Lund, Sweden}
\address[DESYaddress]{Deutsches Elektronen-Synchrotron DESY, Notkestra{\ss}e 85, D-22607 Hamburg, Germany}

\begin{abstract}
In a free-electron laser equipped with variable-gap undulator modules, the technique of undulator tapering opens up the possibility to increase the radiation power beyond the initial saturation point, thus enhancing the efficiency of the laser. The effectiveness of the enhancement relies on the proper optimization of the taper profile. In this work, a multidimensional optimization approach is implemented empirically in the x-ray free-electron laser FLASH2. The empirical results are compared with numerical simulations. 
\end{abstract}


\end{frontmatter}

\section{Introduction}

FLASH~\cite{Schreiber} is the free-electron laser (FEL) facility at the Deutsches Elektronen-Synchrotron (DESY) in Hamburg, Germany. It contains two undulator beamlines, FLASH1 and FLASH2, driven by the same linear accelerator. While FLASH1 consists of fixed-gap undulator modules, FLASH2 is equipped with variable-gap undulator modules. The variable-gap feature enables the simultaneous operation of FLASH1 and FLASH2 at different wavelengths~\citep{Faatz}. It also enables the implementation of undulator tapering in FLASH2.

Undulator tapering involves the variation of the undulator parameter $K$ as a function of the distance $z$ along the undulator line, for the purpose of enhancing the radiation power (and hence the efficiency) of the FEL. This has been demonstrated empirically in x-ray FELs, such as LCLS~\cite{Ratner} and SACLA~\cite{SACLA}. In order to maximize the enhancement of radiation power, the taper profile $K(z)$ needs to be properly optimized.

Present-day imaging experiments at x-ray FELs call for an increased number of photons within a shorter pulse duration~\cite{Neutze,Chapman}. To meet the stringent demand on the radiation power, the theory of taper optimization has been revisited in recent years. In Ref.~\cite{Yurkov}, an important step is made towards the formulation of a universal taper law. In Refs.~\cite{PRSTAB2015, FLASH2_KMR}, taper optimization methods based on the classic Kroll-Morton-Rosenbluth (KMR) model~\cite{KMR} are demonstrated in numerical simulations. In Refs.~\cite{Jiao, Claudio}, a multidimensional optimization method is performed in numerical simulations, whereby the optimal taper profile $K(z)$ is obtained by scanning through a parameter space comprising the taper order (such as linear and quadratic), the taper start point, the taper amplitude etc.

The multidimensional optimization approach is relatively straightforward. Guided by the theoretical studies, this approach is implemented empirically in FLASH2 at a wavelength of 44~nm, and the results are presented in this article. The empirical results of the taper optimization are then compared with the corresponding numerical simulations. The agreement and discrepancies between the empirical and simulation results are analyzed. The article concludes by excluding a number of otherwise possible causes of the discrepancies.

\section{Empirical Study}

\subsection{Machine Parameters}

FLASH2 contains a total of 12 undulator modules. Between every two adjacent modules, there is a drift section for beam focusing, trajectory correction, phase shifting, diagnostics etc. Table~\ref{ExpPars} shows the known machine parameters. For machine parameters not listed in Table~\ref{ExpPars}, the nominal design values~\cite{Schreiber} are assumed. 

\begin{table}[htb]
   \centering
   \begin{tabular}{lcc}
       \hline
       \textbf{Parameter}				& \textbf{Symbol}		& \textbf{Value} \\
       \hline
          Electron beam energy			& $\gamma m_e c^2$		& 646 MeV \\
          Bunch charge					& $Q$					& 300 pC \\
          Radiation wavelength			& $\lambda$				& 44 nm \\
          Repetition rate				& $R$					& 1 MHz \\
          Undulator period 				& $\lambda_w$			& 31.4 mm \\
          Magnetic length per module	& $L_{\text{mod}}$		& 2.5 m \\
          Period of FODO lattice		& $L_{\text{FODO}}$		& 6.6 m \\
       \hline
   \end{tabular}
   \caption{Machine parameters}
   \label{ExpPars}
\end{table}

\subsection{Taper Optimization Scheme}

Each of the 12 undulator modules ($m = 1, 2, ... , 12$) is set to an undulator parameter $K_m$. Within each module, the undulator parameter is uniform. The taper profiles considered in this empirical study are defined by three parameters: the taper order $d$, the start module $n$ and the taper amplitude $\Delta K/K$. These taper profiles are given by the ansatz
\begin{equation}
  K_m = 
  \begin{dcases}
    K			&\text{for $1 \leq m < n$} \\
    K \left[1 - \left( \frac{\Delta K}{K} \right) \left( \frac{m-n+1}{12-n+1} \right)^d \right]
    			&\text{for $n \leq m \leq 12$}.
  \end{dcases}
  \label{TaperProfDeg}
\end{equation}

In Eq.~(\ref{TaperProfDeg}), $K$ is the \textit{initial} undulator parameter, in resonance with the initial energy of the electron beam. The undulator parameter remains $K$ from modules 1 to $n-1$, and decreases in steps from module $n$ onwards. The taper order $d$ equals 1 for linear tapering, and 2 for quadratic tapering. The taper amplitude $\Delta K/K$ is defined such that the undulator parameter of the last module is $K_{12} = K - \Delta K$.

Multidimensional optimization is performed by scanning $d$, $n$ and $\Delta K/K$ empirically for the highest final radiation energy. The same type of multidimensional optimization in numerical simulations is presented in Refs.~\cite{Jiao, Claudio}.

\subsection{Phase Shifter Configuration}
\label{PhaseShifterConfig}

In the drift section between every two undulator modules, there is a phase shifter for the proper matching of the phase between the electron beam and the optical field. The phase shifters are characterized in Ref.~\cite{PhaseShifters}. The required phase shift in each drift section depends solely on the undulator parameter of the preceding undulator module. The phase shifts are implemented automatically by a baseline procedure to ensure constructive interference between the optical fields emitted before and after each drift section. The procedure also accounts for the phase advance caused by the fringe fields at the two ends of each undulator module.

\subsection{Radiation Energy Measurement}

The lasing of FLASH2 takes place through the process of self-amplified spontaneous emission (SASE). The energy of a radiation pulse is measured with a micro-channel plate (MCP) detector~\cite{MCP}, located downstream after the 12 undulator modules. The MCP detector offers a relatively high accuracy over a dynamic range of radiation intensities. To account for the shot-to-shot variability, each energy measurement is averaged over about 100 pulses.

The gas-monitor detector (GMD)~\cite{GMD}, which is also located downstream after the 12 undulator modules, measures the optical pulse energy in parallel. The GMD reading is used as a cross check.

With all the 12 undulator modules engaged, the MCP and GMD measure the \textit{final} pulse energy. To examine the evolution of the pulse energy along the undulator line, it is necessary to measure the \textit{intermediate} pulse energy upstream. To measure the pulse energy immediately after an upstream undulator module, the gaps of all subsequent modules are opened, so that the optical pulse propagates towards the detectors without further interacting with the electron beam. During the propagation, the optical pulse undergoes vacuum diffraction, and its transverse size can increase. So long as the detectors collect the signal of the \textit{entire} optical pulse, the \textit{pulse energy} remains unchanged.

\subsection{Empirical Results}

\begin{figure}[tb]
   \centering
   \includegraphics*[width=\figurewidth]{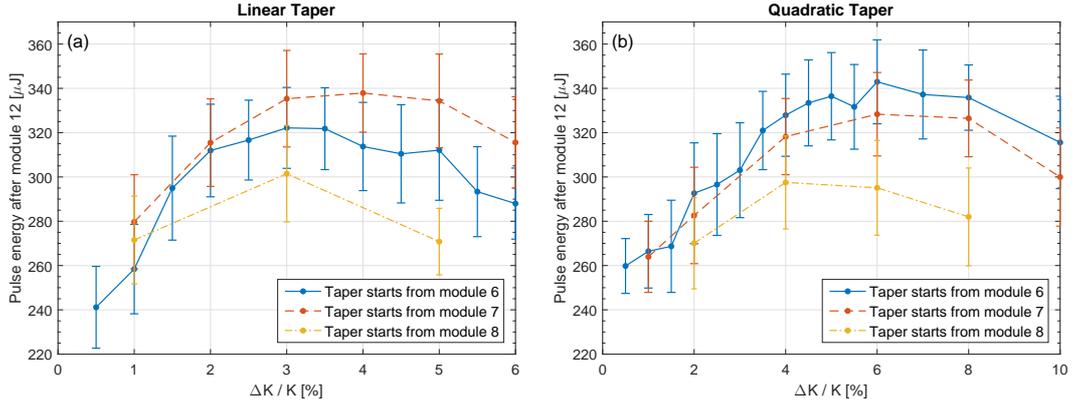}
   \caption{Empirical data. The final pulse energy is plotted as a function of the taper amplitude $\Delta K/K$ for (a) linear tapering ($d = 1$) and (b) quadratic tapering ($d = 2$). The blue solid curve, red dashed curve and yellow dotted curve correspond respectively to start modules $n =$ 6, 7 and 8.}
   \label{DeltaK_plots_exp}
\end{figure}

\begin{figure}[tb]
   \centering
   \includegraphics*[width=\figurewidth]{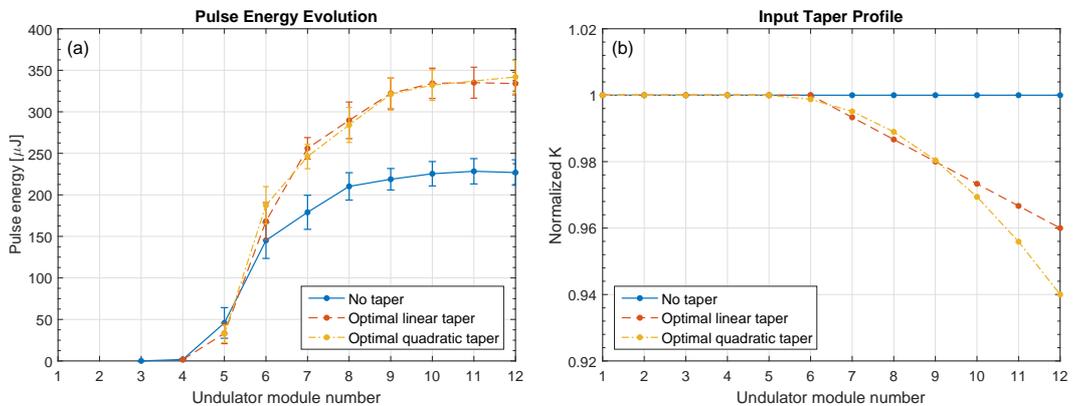}
   \caption{Empirical data. The evolution of the (a) optical pulse energy and (b) input undulator parameter along the undulator line. The undulator parameter is normalized to the initial value. The blue solid curve, red dashed curve and yellow dotted curve correspond respectively to no taper ($\Delta K/K = 0$), the optimal linear taper ($n =7$, $\Delta K/K = 4\%$) and the optimal quadratic taper ($n =6$, $\Delta K/K = 6\%$).}
   \label{Energy_evol_exp}
\end{figure}

The final optical pulse energy is measured for different taper profiles given by Eq.~(\ref{TaperProfDeg}). The measurement is done for taper orders $d = 1, 2$ and for start modules $n = 6, 7, 8$. The results are shown in Fig.~\ref{DeltaK_plots_exp}. Each data point in Fig.~\ref{DeltaK_plots_exp} is obtained with the MCP detector, and is the average over $140 \pm 50$ pulses. The error bar indicates the standard deviation of the MCP readings. Among all the taper profiles considered in Fig.~\ref{DeltaK_plots_exp}, the optimal linear taper occurs at $n = 7$ and $\Delta K/K = 4 \%$, whereas the optimal quadratic taper occurs at $n = 6$ and $\Delta K/K = 6 \%$.

For the optimal linear taper, the optimal quadratic taper and no taper, the intermediate pulse energies are  measured. The evolution of the pulse energy along the undulator line is shown in Fig.~\ref{Energy_evol_exp}(a). The corresponding taper profiles, as input from the control room, are shown in Fig.~\ref{Energy_evol_exp}(b) for reference. Each data point in Fig.~\ref{Energy_evol_exp}(a) is obtained with the MCP detector, and is the average over $110 \pm 30$ pulses. Among all the data points in Figs.~\ref{DeltaK_plots_exp} and~\ref{Energy_evol_exp}(a), the absolute difference between the MCP and GMD values is 19 $\mu$J on average, with a standard deviation of 16 $\mu$J.

In the absence of tapering, the saturation of pulse energy is reached in module 8 [see solid curve in Fig.~\ref{Energy_evol_exp}(a)]. In other words, the start modules ($n = 6,7,8$) considered in Fig.~\ref{DeltaK_plots_exp} are in the vicinity of the initial saturation point.

\section{Comparison with Numerical Simulation}

\subsection{Simulation Parameters}

The empirical results are compared with numerical simulation, after the experiment has been completed. The simulation is performed using the three-dimensional and time-dependent simulation code GENESIS~\cite{GENESIS}, with parameter values as close as possible to the empirical ones (see Table~\ref{ExpPars}). Parameters not specified in Table~\ref{ExpPars} are assumed to have the nominal values shown in Table~\ref{SimPars}.

\begin{table}[htb]
   \centering
   \begin{tabular}{lcc}
       \hline
       \textbf{Parameter}				& \textbf{Symbol}			& \textbf{Value} \\
       \hline
          Peak current					& $I_0$ 					& 1.5 kA \\
          RMS bunch length				& $\sigma_z$				& 24 $\mu$m \\
          RMS energy spread				& $\sigma_\gamma m_e c^2$ 	& 0.5 MeV \\
          Normalized emittance			& $\varepsilon_{x,y}$		& 1.4 mm mrad \\
          Average of beta function		& $\bar{\beta}_{x,y}$		& 6 m \\
       \hline
   \end{tabular}
   \caption{Nominal FLASH2 parameter values used in the simulation}
   \label{SimPars}
\end{table}

In the simulation, the initial values of the optical functions and the quadrupole strengths are chosen self-consistently to give the desired average beta value, independent of the values used in the experiment.

\subsection{Simulation Results}
 
\begin{figure}[tb]
   \centering
   \includegraphics*[width=\figurewidth]{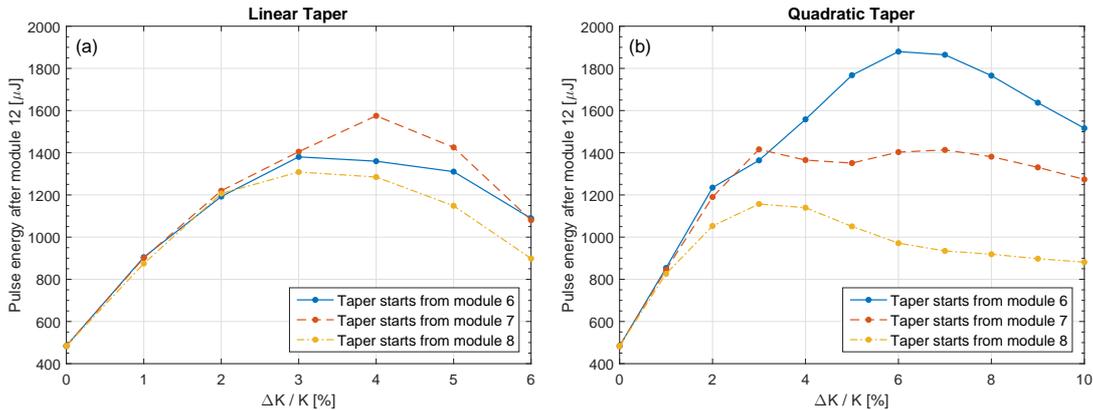}
   \caption{Simulation results. The final pulse energy is plotted as a function of the taper amplitude $\Delta K/K$ for (a) linear tapering ($d = 1$) and (b) quadratic tapering ($d = 2$). The blue solid curve, red dashed curve and yellow dotted curve correspond respectively to start modules $n =$ 6, 7 and 8.}
   \label{DeltaK_plots_sim}
\end{figure}

\begin{figure}[tb]
   \centering
   \includegraphics*[width=\figurewidth]{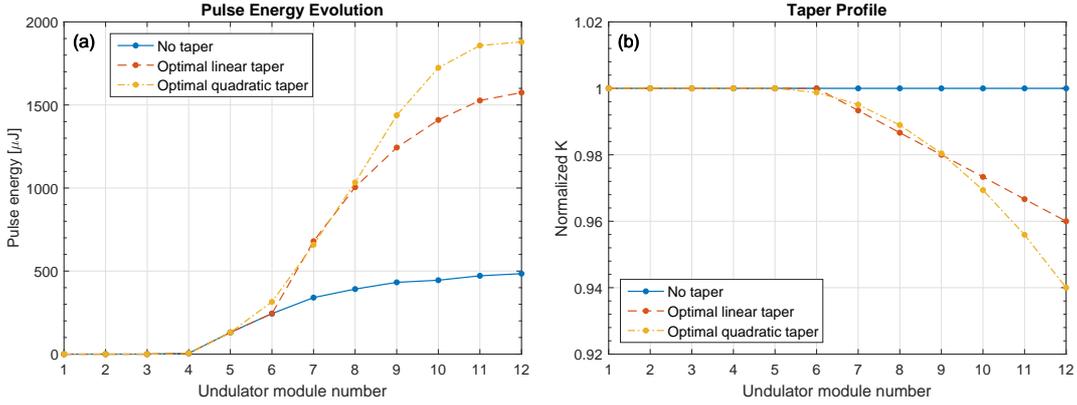}
   \caption{Simulation results. The evolution of the (a) optical pulse energy and (b) undulator parameter along the undulator line. The undulator parameter is normalized to the initial value. The blue solid curve, red dashed curve and yellow dotted curve correspond respectively to no taper ($\Delta K/K = 0$), the optimal linear taper ($n =7$, $\Delta K/K = 4\%$) and the optimal quadratic taper ($n =6$, $\Delta K/K = 6\%$).}
   \label{Energy_evol_sim}
\end{figure}

The same multidimensional optimization is performed in simulation. Using Eq.~(\ref{TaperProfDeg}) as the ansatz, the parameters $d$, $n$ and $\Delta K/K$ are scanned for the highest final radiation energy. The results are shown in Fig.~\ref{DeltaK_plots_sim}. Among all the taper profiles considered in Fig.~\ref{DeltaK_plots_sim}, the optimal linear taper occurs at $n = 7$ and $\Delta K/K = 4 \%$, whereas the optimal quadratic taper occurs at $n = 6$ and $\Delta K/K = 6 \%$.

For the optimal linear taper, the optimal quadratic taper and no taper, the simulated pulse energy evolutions along the undulator line are shown in Fig.~\ref{Energy_evol_sim}(a). The corresponding taper profiles are shown in Fig.~\ref{Energy_evol_sim}(b) for reference.

\subsection{Comparing Empirical and Simulation Results}

Comparing Fig.~\ref{DeltaK_plots_exp} (empirical) and Fig.~\ref{DeltaK_plots_sim} (simulation), the optimal taper profiles are consistent. In both cases, the optimal linear taper occurs at $n = 7$ and $\Delta K/K = 4 \%$, and the optimal quadratic taper occurs at $n = 6$ and $\Delta K/K = 6 \%$. Figs.~\ref{DeltaK_plots_exp} and~\ref{DeltaK_plots_sim} also show good agreement in the overall trend for the final optical pulse energy $\mathcal{E}$. In both cases, the overall trend for linear tapering ($d = 1$) is
\begin{equation*}
\mathcal{E} (n=7, \Delta K/K) > \mathcal{E} (n=6, \Delta K/K) > \mathcal{E} (n=8, \Delta K/K),
\end{equation*}
whereas the overall trend for quadratic tapering ($d = 2$) is
\begin{equation*}
\mathcal{E} (n=6, \Delta K/K) > \mathcal{E} (n=7, \Delta K/K) > \mathcal{E} (n=8, \Delta K/K).
\end{equation*}
However, Figs.~\ref{DeltaK_plots_exp} and~\ref{DeltaK_plots_sim} show disagreement in terms of the absolute pulse energies. The range of pulse energies is generally higher in the simulation than in the experiment.

Next, the pulse energy evolution along the undulator line is compared between simulation [see Fig.~\ref{Energy_evol_sim}(a)] and experiment [see Fig.~\ref{Energy_evol_exp}(a)]. In both cases, the pulse energy remains in the order of 1 $\mu$J before module 5, and exceeds the 10-$\mu$J threshold in module 5. In the absence of tapering, the initial saturation point is situated around module 8 in both cases (see solid curves). With the optimal linear and quadratic tapers, final saturation is reached within the 12 undulator modules in both simulation and experiment, but occurs earlier in the experiment than in the simulation (see dashed and dotted curves).

In the experiment, the optimal linear taper and the optimal quadratic taper yield almost identical final pulse energy. But in the simulation, the final pulse energy for the optimal quadratic taper is 1.2 times higher than that for the optimal linear taper.

In the experiment, the enhancement factor is 
\begin{equation*}
\frac{\mathcal{E} (\text{optimal taper})}{\mathcal{E} (\text{no taper})} = 1.5.
\end{equation*}
But in the simulation, the enhancement factor is
\begin{equation*}
\frac{\mathcal{E} (\text{optimal taper})}{\mathcal{E} (\text{no taper})} = 3.9,
\end{equation*}
which is 2.6 times higher than that in the experiment.

\subsection{Discussion on the Taper Start Point}

In both the simulation and empirical results, the optimal linear taper starts from module 7, while the optimal quadratic taper starts from module 6. The reason for this difference in the optimal start point is that the undulator parameter decreases much more slowly at the beginning of the quadratic taper. This is seen in Figs.~\ref{Energy_evol_exp}(b) and~\ref{Energy_evol_sim}(b). In module 6 from which the quadratic taper starts, the undulator parameter $K_6$ is effectively identical to the initial value $K_1$, as $K_6 = 99.88\% \times K_1 \approx K_1$. It is in module 7 where the undulator parameter starts to show a significant difference from the initial value. In other words, the optimal quadratic taper starts effectively from module 7, the same module from which the optimal linear taper starts.

Refs.~\cite{Yurkov, Fawley} suggest that the optimal taper start point is two gain lengths before the initial saturation point. In one-dimensional theory, the gain length is given by
\begin{equation}
L_g = \frac{\lambda_w}{4 \sqrt{3} \pi \rho},
\label{Lg}
\end{equation}
where
\begin{equation}
\rho = \frac{1}{4} \left( \frac{I}{I_A \varepsilon_{x,y} \bar{\beta}_{x,y}} \right)^{1/3}
    \left( \frac{\lambda_w K f_B}{\pi \gamma} \right)^{2/3}
\label{Pierce}
\end{equation}
is the dimensionless Pierce parameter, $I_A = m_e c^3 /e = 17.045$ kA is the Alfv\'{e}n current, $\sigma_x$ is the rms radius of the electron beam, and  $f_B = J_0 (\xi) - J_1 (\xi)$ is the Bessel factor for planar undulators, with $\xi = K^2 / [2(K^2+2)]$.

With the parameters in Tables~\ref{ExpPars} and~\ref{SimPars}, the Pierce parameter is $\rho = 3.51 \times 10^{-3}$, and the gain length is $L_g = 0.41$ m. Thus, the optimal taper start point is predicted to be $2 L_g = 0.82$ m before the initial saturation point, excluding the length of the drift section between undulator modules. If we assume that the precise initial saturation point is at the beginning of module 8, then the optimal taper start point should lie within module 7. This rough prediction agrees with the simulation and empirical results.

\subsection{Relating Optimal Taper Profiles to the KMR Model}

\begin{figure}[tb]
   \centering
   \includegraphics*[width=\figurewidth]{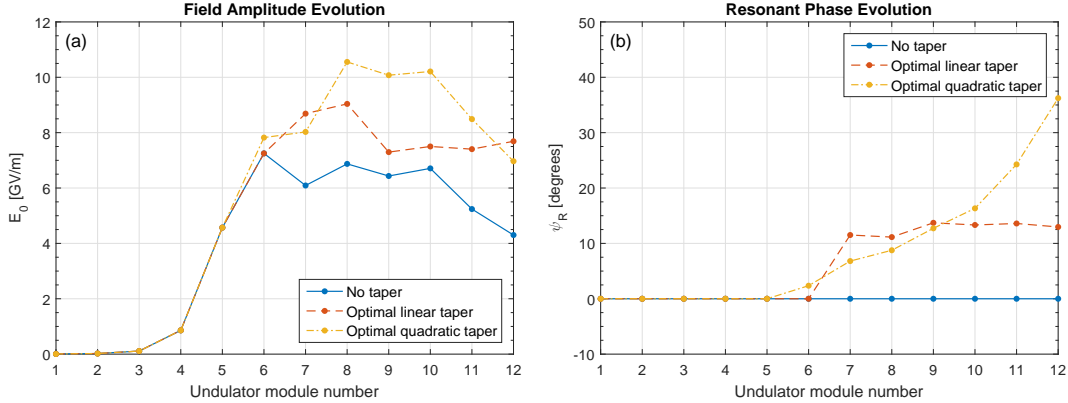}
   \caption{Simulation results. The evolution of the (a) optical field amplitude $E_0$ and (b) resonant phase $\psi_R$ along the undulator line. The blue solid curve, red dashed curve and yellow dotted curve correspond respectively to no taper ($\Delta K/K = 0$), the optimal linear taper ($n =7$, $\Delta K/K = 4\%$) and the optimal quadratic taper ($n =6$, $\Delta K/K = 6\%$).}
   \label{psiR_evol_sim}
\end{figure}

The Kroll-Morton-Rosenbluth (KMR) model~\cite{KMR} is a theoretical analysis of undulator tapering in FELs based on a one-dimensional relativistic Hamiltonian formulation. In Refs.~\cite{PRSTAB2015, FLASH2_KMR}, the KMR model is used as a method to optimize FEL taper profiles in numerical simulations. After choosing the resonant phase $\psi_R (z)$, the taper profile $K(z)$ is computed from the differential equation
\begin{equation}
\label{KMR_equation}
\frac{dK}{dz} = - \frac{2 e}{m_e c^2} \frac{\lambda}{\lambda_w} f_B (z) E_0 (z) \sin[\psi_R (z)],
\end{equation}
where $E_0$ is the on-axis field amplitude and $z$ is the position along the undulator line. With a constant $\psi_R$, the optimization is known as the ordinary KMR method. With a variable $\psi_R$ which increases gradually from zero, the optimization is known as the modified KMR method.

With the simulation results at hand, the evolution of the resonant phase $\psi_R$ along the undulator line can be back-calculated from Eq.~(\ref{KMR_equation}). This back-calculation requires the taper profile $K(z)$ [see Fig.~\ref{Energy_evol_sim}(b)] and the field amplitude evolution $E_0 (z)$ [see Fig.~\ref{psiR_evol_sim}(a)] as inputs. Carrying out this back-calculation for the optimal linear taper, the optimal quadratic taper and no taper, the resulting $\psi_R (z)$ functions are shown in Fig.~\ref{psiR_evol_sim}(b).

The optimal linear and quadratic tapers start from module 7 and module 6, respectively. Before the taper starts, $\psi_R = 0$ [see dashed and dotted curves in Fig.~\ref{psiR_evol_sim}(b)]. This is expected, as $dK/dz = 0$ implies $\psi_R = 0$ according to Eq.~(\ref{KMR_equation}). For the same reason, in the absence of any tapering, $\psi_R$ remains zero at all times [see solid line in Fig.~\ref{psiR_evol_sim}(b)].

When the optimal linear taper starts in module 7, $\psi_R$ increases abruptly from 0 to $12^{\circ}$, and remains almost constant afterwards [see dashed curve in Fig.~\ref{psiR_evol_sim}(b)]. When the optimal quadratic taper starts in module 6, $\psi_R$ increases gradually and monotonically from 0, until it reaches a value of $36^{\circ}$ in the final module [see dotted curve in Fig.~\ref{psiR_evol_sim}(b)]. The $\psi_R (z)$ function for the optimal linear taper resembles one used in the ordinary KMR method, whereas the $\psi_R (z)$ function for the optimal quadratic taper resembles one used in the modified KMR method.

\section{Post-Experimental Analysis of the Discrepancies}

\subsection{General Remarks}

The empirical and simulation results are in good agreement in terms of:
\begin{itemize}
\item the $(n, \Delta K/K)$ values for the optimal linear and quadratic tapers;
\item the overall trend in the plots of the final energy $\mathcal{E}$ versus $\Delta K/K$ (see Figs.~\ref{DeltaK_plots_exp} and~\ref{DeltaK_plots_sim}); and
\item the module in which the exponential gain crosses the 10-$\mu$J threshold (see Figs.~\ref{Energy_evol_exp} and~\ref{Energy_evol_sim}).
\end{itemize}
However, there are three main discrepancies between the empirical and simulation results:
\begin{itemize}
\item In the parameter space $(d,n,\Delta K/K)$ considered, the $\mathcal{E}$ range is generally lower in the experiment than in the simulation (see Figs.~\ref{DeltaK_plots_exp} and~\ref{DeltaK_plots_sim}).
\item The enhancement factor $\mathcal{E} (\text{optimal taper})/\mathcal{E} (\text{no taper})$ is 3.9 in the simulation, but only 1.5 in the experiment.
\item With the optimal linear and quadratic tapers, final saturation occurs earlier in the experiment than in the simulation (see Figs.~\ref{Energy_evol_exp} and~\ref{Energy_evol_sim}). 
\end{itemize}

The exact causes of these discrepancies are not known. Yet, it is possible to \textit{exclude} a number of otherwise possible causes, such as the shot-to-shot variability, drift of the machine and wakefield effects. These are addressed in the upcoming subsections.

The discrepancies in question can also be caused by incorrect assumptions of parameter values. For the simulation, the nominal FLASH2 parameter values in Table~\ref{SimPars} are assumed. The assumed nominal values in the simulation can be different from the unknown actual values in the experiment.

As illustrated in the sensitivity study in Ref.~\cite{FEL2014}, a slight change in the emittance, energy spread or peak current can have a huge impact on the optimized radiation power of a tapered FEL. In other words, if the actual emittance, energy spread or peak current is worse than assumed, then the optimized radiation energy will be lower than expected. This will, in turn, influence the enhancement factor. This can possibly explain the discrepancies in question. However, the proposition that the emittance, energy spread or peak current is worse than assumed will be \textit{disproved} in the following subsections.

\subsection{Shot-to-Shot Variability}

In the empirical results (Figs.~\ref{DeltaK_plots_exp} and~\ref{Energy_evol_exp}), the shot-to-shot fluctuations are accounted for by the error bar, which indicates the standard deviation of many shots. All the error bars are within $\pm$ 23 $\mu$J, which is too small to account for the discrepancies between the simulation and empirical results.

\subsection{Drift of the Machine}

Consider two scenarios in particular, the optimal linear taper and no taper. Since the optimal linear taper only starts from undulator module 7, the two scenarios are identical before module 7. In principle, the two scenarios should yield the same pulse energy evolution before module 7. This is precisely the case in the simulation [see solid and dashed curves in Fig.~\ref{Energy_evol_sim}(a)], which is the ideal case free of any drift. But in the empirical results [see solid and dashed curves in Fig.~\ref{Energy_evol_exp}(a)], the two scenarios yield slightly different energies in modules 5 and~6. The energy differences can be partly attributed to the drift of the machine. But despite the drift, the energy differences are still within 24 $\mu$J, which is too small to account for the discrepancies between the simulation and empirical results.

\subsection{Emittance Underestimated}
\label{EmittanceUnderestimated}

\begin{figure}[tb]
   \centering
   \includegraphics*[width=\figurewidth]{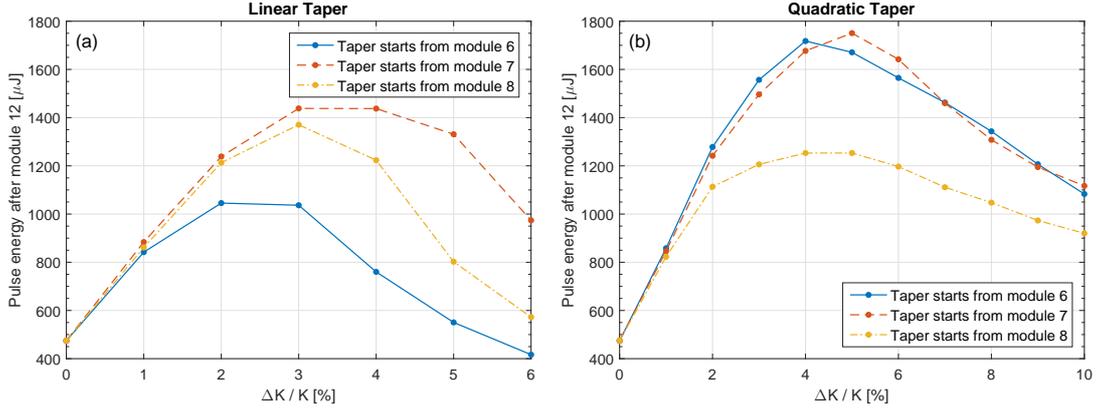}
   \caption{Simulation results with normalized emittance increased from 1.4 mm mrad to 1.6 mm mrad. The final pulse energy is plotted as a function of the taper amplitude $\Delta K/K$ for (a) linear tapering and (b) quadratic tapering.}
   \label{DeltaK_plots_sim_NormEmit}
\end{figure}
In order to disprove that the emittance is underestimated, the simulation is repeated with the normalized emittance  slightly increased, from 1.4 mm mrad to 1.6 mm mrad. All other parameters in Tables~\ref{ExpPars} and~\ref{SimPars} are kept unchanged. With the average beta function $\bar{\beta}_{x,y}$ kept unchanged, this requires increasing the RMS beam radius $\sigma_{x,y}$ from 82 $\mu$m to 87 $\mu$m. The new simulation results are shown in Fig.~\ref{DeltaK_plots_sim_NormEmit}.

If the emittance were indeed underestimated in the original simulation, then the new simulation (with an increased emittance) would show an improved agreement with the empirical results. But in the new simulation results, the overall trends of the final pulse energy $\mathcal{E}$ change. As seen in Fig.~\ref{DeltaK_plots_sim_NormEmit}, the overall trend for linear tapering ($d = 1$) becomes
\begin{equation*}
\mathcal{E} (n=7, \Delta K/K) > \mathcal{E} (n=8, \Delta K/K) > \mathcal{E} (n=6, \Delta K/K),
\end{equation*}
whereas the overall trend for quadratic tapering ($d = 2$) becomes
\begin{equation*}
\mathcal{E} (n=6, \Delta K/K) \approx \mathcal{E} (n=7, \Delta K/K) > \mathcal{E} (n=8, \Delta K/K).
\end{equation*}
The overall trends actually become further off from those in the empirical results (see Fig.~\ref{DeltaK_plots_exp}). Meanwhile, there is no improved agreement in the $\mathcal{E}$ range and in the enhancement factor. This disproves that the emittance is underestimated in the original simulation.

Comparing the two sets of simulation results in Fig.~\ref{DeltaK_plots_sim} and~\ref{DeltaK_plots_sim_NormEmit}, the increased emittance makes it more favourable to start the taper at a later point down the undulator line. The optimal quadratic taper in Fig.~\ref{DeltaK_plots_sim} starts from module 6, whereas that in Fig.~\ref{DeltaK_plots_sim_NormEmit} starts from module 7. As for linear taper, module 7 remains the most favourable start module. Yet, while module 8 is the least favourable of the three start modules considered in Fig.~\ref{DeltaK_plots_sim}, it becomes the second most favourable in Fig.~\ref{DeltaK_plots_sim_NormEmit}.

The shift in the optimal taper start point can be explained as follows. Refs.~\cite{Yurkov, Fawley} suggest that the optimal taper start point $z_0$ is two gain lengths before the initial saturation point. In one-dimensional theory, this is given by
\begin{equation}
z_0 = L_{\text{sat}} - 2 L_g = \frac{\lambda_w}{\rho} - 2 \left( \frac{\lambda_w}{4 \sqrt{3} \pi \rho} \right) = 
\left( 1 - \frac{1}{2 \sqrt{3} \pi} \right) \frac{\lambda_w}{\rho} \propto \frac{1}{\rho}.
\label{z0}
\end{equation}
With the definition of the Pierce parameter in Eq.~(\ref{Pierce}), one can deduce that
\begin{equation}
z_0 \propto \frac{1}{\rho} \propto (\varepsilon_{x,y})^{1/3}.
\label{z0_emit}
\end{equation}
The proportionality implies that an increased emittance moves the optimal taper start point downstream. It also implies that a further increase in emittance would move the optimal taper start point further downstream, thus making the overall trends of $\mathcal{E}$ even further off from those in the empirical results.

\subsection{Peak Current Overestimated}
\label{PeakCurrentOverestimated}

\begin{figure}[tb]
   \centering
   \includegraphics*[width=\figurewidth]{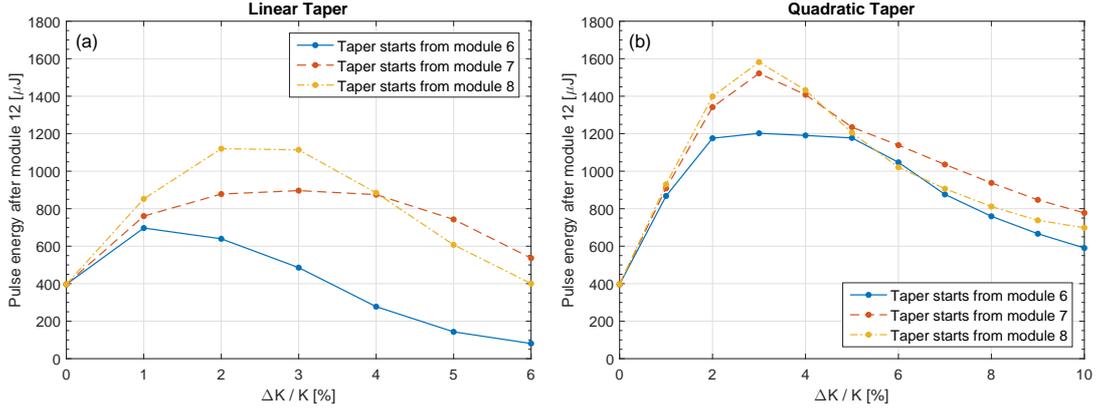}
   \caption{Simulation results with peak current decreased from 1.5 kA to 1.2 kA. The final pulse energy is plotted as a function of the taper amplitude $\Delta K/K$ for (a) linear tapering and (b) quadratic tapering.}
   \label{DeltaK_plots_sim_PCurrent}
\end{figure}

In order to disprove that the peak current is overestimated, the simulation is repeated with the peak current slightly decreased, from 1.5 kA to 1.2 kA. In order to keep the known bunch charge in Table~\ref{ExpPars} unchanged, this requires increasing the RMS bunch length from 24 $\mu$m to 30 $\mu$m. All other parameters in Tables~\ref{ExpPars} and~\ref{SimPars} are kept unchanged. The new simulation results are shown in Fig.~\ref{DeltaK_plots_sim_PCurrent}.

Again, by decreasing the peak current in the simulation, the overall trends in the final pulse energy $\mathcal{E}$ become further off from those in the empirical results (see Fig.~\ref{DeltaK_plots_exp}). This disproves that the peak current is overestimated.

Comparing the two sets of simulation results in Fig.~\ref{DeltaK_plots_sim} and~\ref{DeltaK_plots_sim_PCurrent}, the decreased peak current also makes it more favourable to start the taper in a later undulator module. This agrees with the one-dimensional theoretical prediction from Eqs.~(\ref{Pierce}) and~(\ref{z0}) that
\begin{equation}
z_0 \propto \frac{1}{\rho} \propto \frac{1}{I^{1/3}}.
\label{z0_current}
\end{equation}

\subsection{Energy Spread Underestimated}
\label{EnergySpreadUnderestimated}

\begin{figure}[tb]
   \centering
   \includegraphics*[width=\figurewidth]{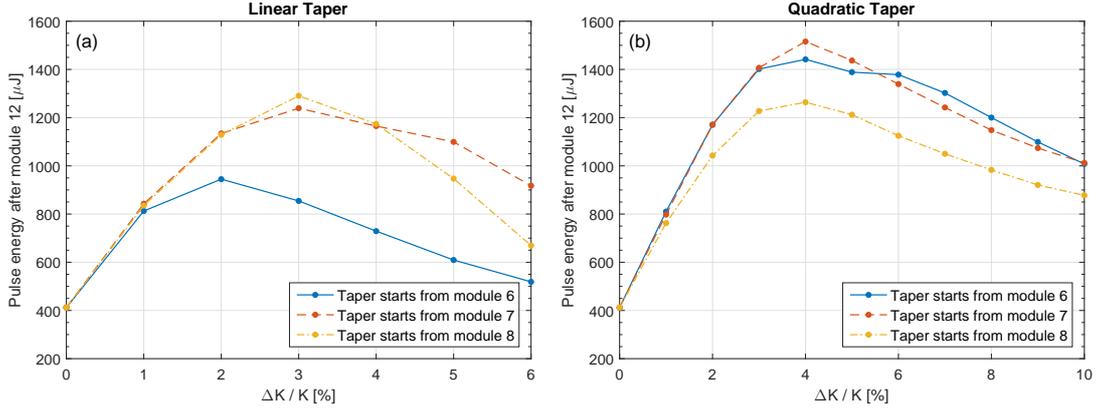}
   \caption{Simulation results with energy spread increased from 0.5 MeV to 0.7 MeV. The final pulse energy is plotted as a function of the taper amplitude $\Delta K/K$ for (a) linear tapering and (b) quadratic tapering.}
   \label{DeltaK_plots_sim_ESpread}
\end{figure}

In order to disprove that the energy spread is underestimated, the simulation is repeated with the energy spread slightly increased, from 0.5 MeV to 0.7 MeV. All other parameters in Tables~\ref{ExpPars} and~\ref{SimPars} are kept unchanged. The new simulation results are shown in Fig.~\ref{DeltaK_plots_sim_ESpread}.

Again, by increasing the energy spread in the simulation, the overall trends in the final pulse energy $\mathcal{E}$ become further off from those in the empirical results (see Fig.~\ref{DeltaK_plots_exp}). This disproves that the energy spread is underestimated.

Comparing the two sets of simulation results in Fig.~\ref{DeltaK_plots_sim} and~\ref{DeltaK_plots_sim_ESpread}, the increased energy spread also makes it more favourable to start the taper in a later undulator module. However, it is impossible to use the one-dimensional formulation to explain the shift in the optimal taper start point caused by the increased energy spread, as it is done for the emittance and the peak current. Nonetheless, the energy spread effects can be explained by similar arguments using the generalized formulation of Ming Xie~\cite{Ming}.

\subsection{Wakefield Effects}

In Ref.~\cite{GENESIS}, a simulation study on the effects of wakefields is performed on a case of the TTF-FEL, which is the predecessor of the FLASH1 and FLASH2 facilities. The machine parameters used in the simulation study are in the same orders of magnitude as those in Tables~\ref{ExpPars} and~\ref{SimPars}. The study identifies three major sources of wakefields, namely, the conductivity, surface roughness and geometrical changes of the beam pipe along the undulator. The simulation on the TTF-FEL case shows that wakefields can reduce the saturation power of the FEL by three orders of magnitude, while keeping the saturation length almost unchanged. In principle, wakefield effects can be a possible explanation for the discrepancies between our empirical and simulation results for FLASH2. However, this can be disproved as follows.

In the empirical optimization of undulator tapering, the optimal taper profile which maximizes the final radiation energy is also that which best compensates the energy loss due to wakefields~\cite{SACLA}.  Meanwhile, in the simulation which results in Fig.~\ref{DeltaK_plots_sim}, wakefields are \textit{not} considered. If wakefield effects were significant, then the optimal taper profile should occur at very different $(n, \Delta K/K)$ values in the empirical and simulation results. But as seen in Figs.~\ref{DeltaK_plots_exp} and~\ref{DeltaK_plots_sim}, this is \textit{not} the case. In fact, the experiment and simulation yield the exact same $(n, \Delta K/K)$ values for the optimal linear taper, and for the optimal quadratic taper. This leads us to the conclusion that wakefield effects are not significant in the experiment, and therefore do not account for the discrepancies in question.

\subsection{Beam Trajectory Errors}
\label{TrajErr}

The ideal trajectory of the electron beam is the central axis along the undulator line. But if the electron beam undergoes betatron oscillations as a whole, it deviates from the ideal trajectory and is subject to trajectory errors. These errors can be caused by a combination of many factors, which include
\begin{itemize}
\item the imperfect alignment of the undulator modules;
\item the imperfect alignment of the quadrupole magnets; and
\item the inclined injection of the electron beam to the undulator modules.
\end{itemize}

Trajectory errors can degrade the FEL performance through a number of mechanisms~\cite{Tanaka}. A complete analysis of all these mechanisms is not trivial. But in taper optimization studies, it is the undulator parameter $K$ which characterizes a taper profile. The following discussions shall focus on the implication of trajectory errors to $K$. 

The undulator parameter $K$ is associated with the magnetic field strength $B_0$ on the central axis of the undulator by the definition
\begin{equation}
\label{K_def}
K = \frac{e \lambda_w}{2 \pi m_e c}  B_0.
\end{equation}
In the presence of trajectory errors, the electron beam deviates from the central axis. Even if the on-axis field strength $B_0$ were perfectly accurate, the electron beam would still experience a field strength different from the desired value $B_0$, hence an undulator parameter different from the desired value $K$. As derived in the Appendix, the effective undulator parameter is 
\begin{equation}
K_\textrm{eff} = K \cosh (k_w y) \geq K,
\label{Keff}
\end{equation}
where $y$ is the deviation of the electron beam from the central axis, and $k_w = 2 \pi / \lambda_w$ is the undulator wavenumber. The magnetic field strength experienced by an electron beam with a trajectory error $y$ in an undulator with parameter $K$ is equivalent to that experienced by an on-axis electron beam in an undulator of parameter $K_\textrm{eff}$.

The effective undulator parameter $K_\textrm{eff}$ also leads to a phase shift error. As mentioned in Section~\ref{PhaseShifterConfig}, the required phase shift in the drift section depends solely on the $K$ value of the preceding undulator module. Given an input value $K$, the phase shifter is automatically adjusted to ensure proper phase matching at the end of the drift section. But if the effective value is $K_\textrm{eff} \neq K$, then a phase mismatch will occur. As derived in the Appendix, this phase mismatch is given by
\begin{equation}
\delta \phi = - k_w L_D K^2 \left( 1 + \frac{K^2}{2} \right)^{-2} [\cosh(k_w y) - 1],
\label{delphi}
\end{equation}
Here $L_D$ is the drift section length, which is 800 mm in FLASH2.

\begin{figure}[tb]
   \centering
   \includegraphics*[width=\figurewidth]{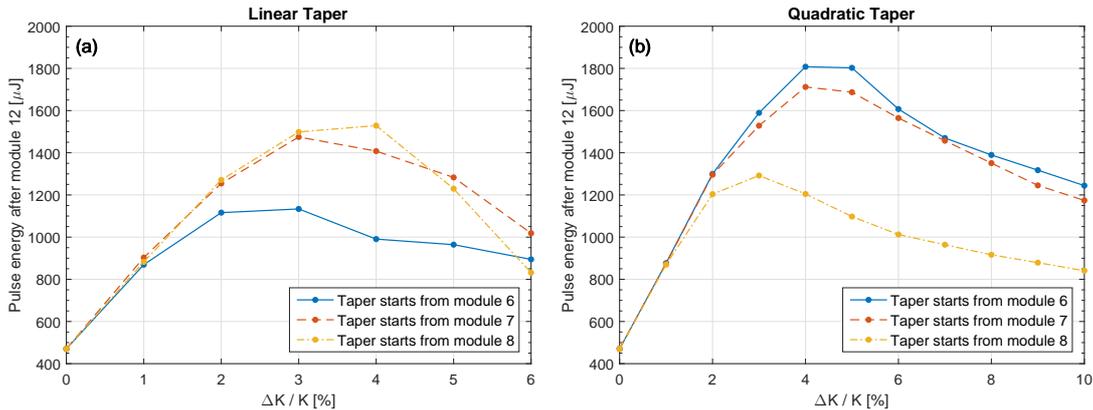}
   \caption{Simulation results for $K_{\text{eff}}$ and $\delta\phi$ associated with a trajectory error of $y = 250$ $\mu$m. The final pulse energy is plotted as a function of the taper amplitude $\Delta K/K$ for (a) linear tapering and (b) quadratic tapering.}
   \label{DeltaK_plots_sim_OrbitErr}
\end{figure}

The simulation is now repeated with the $K_{\text{eff}}$ and $\delta\phi$ associated with a trajectory error of $y = 250$ $\mu$m, calculated from Eqs.~(\ref{Keff}) and~(\ref{delphi}). With a trajectory error of $y = 250$ $\mu$m, the difference between $K_\textrm{eff}$ and $K$ becomes comparable to the Pierce parameter $\rho$, and is therefore  significant. The new simulation results are shown in Fig.~\ref{DeltaK_plots_sim_OrbitErr}. Again, the overall trends in the final pulse energy $\mathcal{E}$ become further off from those in the empirical results (see Fig.~\ref{DeltaK_plots_exp}). Thus, the $K_{\text{eff}}$ and $\delta\phi$ associated of a trajectory error of $y = 250$ $\mu$m cannot account for the discrepancies between the empirical and simulation results.

\subsection{Combination of Different Factors}

In the preceding discussions, the different possible causes of the discrepancies in question are considered separately. In the following, combinations of these factors will be discussed.

Shot-to-shot fluctuations and the drift of the machine can each affect the measured optical pulse energy by about 20 $\mu$J. The combined effect is then 40 $\mu$J, which is still too small to account for the discrepancies in question.

The emittance, the peak current and the energy spread have been considered individually. As discussed in Sections~\ref{EmittanceUnderestimated}--\ref{EnergySpreadUnderestimated}, if any of these three parameters is worse than assumed, then the optimal taper start point $z_0$ will be shifted downstream [see e.g. Eqs.~(\ref{z0_emit}) and~(\ref{z0_current})]. From this one can deduce that if all three (or at least two of the three) parameters are worse than assumed, then the optical taper start point $z_0$ will be shifted even further downstream. This will, in turn, make the overall trends of the final pulse energy $\mathcal{E}$ even further off from those in the empirical results. Thus, the discrepancies between the simulation and empirical results cannot be explained by the combination of an underestimated emittance, an overestimated peak current and an underestimated energy spread.

There are no indications that the three parameters are much different from their design values. But in principle, one could consider different scenarios where one parameter is worse than assumed while another parameter is \textit{better} than assumed. One example examined in numerical simulation is the scenario where the normalized emittance is halved while the energy spread is doubled (results not shown). The resulting range of optical pulse energies becomes closer to that in the experiment. Yet, the overall trends of the final pulse energy $\mathcal{E}$, as well as the $(n, \Delta K/K)$ values of the optimal tapers, become further off from those in the experiment.

Even though there are possible explanations for some of the discrepancies between the empirical and simulation results, there is no simple explanation that would explain all differences.

\section{Conclusion}

A multidimensional optimization method has been implemented empirically in FLASH2, to optimize the taper profile for the maximum radiation energy. The empirical results have been correlated to simulations.

In the empirical study, the taper profile is characterized by the taper order $d$, the start module $n$ and the taper amplitude $\Delta K/K$. For the optimal linear ($d=1$) and quadratic ($d=2$) tapers, the evolution of the optical pulse energy along the undulator line was examined.

The empirical results were compared with the corresponding results of numerical simulation. The two sets of results show good agreement in terms of the overall trend in the variation of the final pulse energy $\mathcal{E}$ with $\Delta K/K$. They also show good agreement for the optimal linear and quadratic tapers regarding the start module ($n$), the taper amplitude ($\Delta K/K$) and the exponential gain profile. However, there are discrepancies in terms of the general range of pulse energies, the enhancement factor from tapering, as well as the final saturation points for the optimal tapers.

Possible causes of the discrepancies have been examined, and a number of them excluded, such as emittance, energy spread and peak current deviations. Also, shot-to-shot variation, the drift of the machine, wakefield effects, as well as the systematic $K$ and phase shift errors associated with a beam trajectory error have been excluded.

Remaining factors are mainly (i) a poor overlapping between the electron beam and the optical mode, caused by the misalignment and mismatch of the electron optics; and (ii) phase mismatch caused by random errors in the phase shifters. These remaining factors need to be investigated in more detail. Further studies in numerical simulations and empirical measurements are planned for the future.

\section*{Acknowledgment}

The authors would like to thank Katja Honkavaara and Siegfried Schreiber for their crucial roles in facilitating this international collaboration. The authors would also like to thank Evgeny Schneidmiller, Markus Tischer and Mikhail Yurkov for their participation in the planning meeting for the experimental work.

\section*{Appendix: Derivation of $K_\textrm{eff}$ and $\delta\phi$}

In Section~\ref{TrajErr}, the effective undulator parameter $K_\textrm{eff}$ associated with a trajectory error is discussed. This Appendix gives a derivation for $K_\textrm{eff}$ and the subsequent phase mismatch $\delta\phi$. 

Consider a pair of magnetic poles in the undulator, directly opposite to each other. Define the $y$-axis as the straight line passing through the middle points of the two pole tips. As usual, the $z$-axis is in the direction of beam propagation, perpendicular to the $y$-axis. A trajectory error in the $y$-direction changes the distance between the electron beam and the magnetic pole, which has a strong impact on the magnetic field strength experienced by the beam. Meanwhile, an trajectory error purely in the $x$-direction imposes no change on the beam-pole distance, and is therefore not treated here. 

Following the derivation in Ref.~\cite{Schmueser}, the variation of the magnetic field strength $B_y$ along the $y$-axis is examined using a two-dimensional model in the $yz$-plane. Along the $z$-axis, the magnetic field strength is periodic, with a period of $\lambda_w = 2 \pi / k_w$. Assuming that the periodic variation is perfectly sinusoidal, the following ansatz can be written for the magnetic scalar potential:
\begin{equation}
\label{Eq:scalarpotential}
\varphi (y, z) = f(y) \cos(k_w z).
\end{equation}
Here $f (y)$ is an unknown function which depends only on $y$. The scalar potential $\varphi$ has to satisfy the Laplace equation
\begin{equation}
\label{Eq:Laplace}
\nabla^2 \varphi (y, z) = 0.
\end{equation}
Substituting Eq.~(\ref{Eq:scalarpotential}) into Eq.~(\ref{Eq:Laplace}) results in the second-order ordinary differential equation
\begin{equation}
\label{Eq:Laplace2}
\frac{d^2 f(y)}{dy^2} -k_w^2 f(y) = 0,
\end{equation}
to which the general solution is
\begin{equation}
\label{Eq:GenSol}
f(y) = A_1 \sinh (k_w y) + A_2 \cosh (k_w y) 
\end{equation}
with arbitrary constants $A_1$ and $A_2$. Inserting this into Eq.~(\ref{Eq:scalarpotential}), the scalar potential may be rewritten as
\begin{equation}
\label{Eq:scalarpotential2}
\varphi (y, z) = A_1 \sinh (k_w y) \cos(k_w z) + A_2 \cosh (k_w y) \cos(k_w z).
\end{equation}
The $y$-component of the magnetic field is then
\begin{equation}
\label{Eq:By}
B_y (y, z) = - \frac{\partial \varphi}{\partial y} = - k_w A_1 \cosh (k_w y) \cos(k_w z) - k_w A_2 \sinh (k_w y) \cos(k_w z).
\end{equation}

Recalling that the peak field on the $z$-axis is $B_y (0,0) =  B_0$, we have $A_1 = - B_0 / k_w$. Given the symmetry of the system about the plane $y=0$, we have $B_y (+y,z) = B_y (-y,z)$ and hence $A_2 = 0$. With these results, Eq.~(\ref{Eq:By}) can be rewritten as
\begin{equation}
\label{Eq:By2}
B_y (y, z) = B_0 \cosh (k_w y) \cos(k_w z).
\end{equation}
To examine the variation of $B_y$ along the $y$-axis, we set $z$ = 0 and obtain
\begin{equation}
\label{Eq:By_along_y}
B_y (y, 0) = B_0 \cosh (k_w y).
\end{equation}

\begin{figure}[tb]
   \centering
   \includegraphics*[width=\figurewidth]{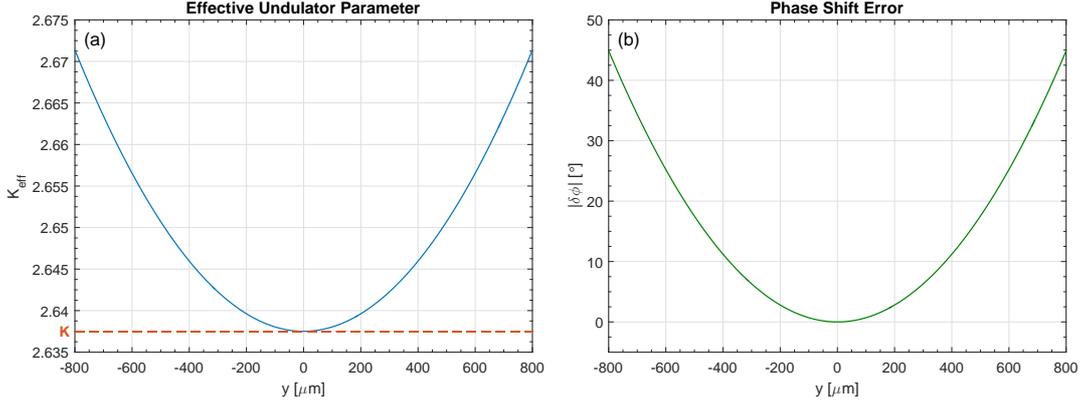}
   \caption{The following quantities are plotted as functions of the trajectory error $y$: (a) the effective undulator parameter $K_{\text{eff}}$ of an undulator module and (b) the resulting error $|\delta\phi|$ in the phase shift immediately after the undulator module. These plots are made for a desired $K$ value of 2.638, which is in resonance with the initial energy of the electron beam.}
   \label{Keff_vs_y}
\end{figure}

In other words, if the electron beam has a trajectory error of $y$, then it experiences a field $B_y (y, 0)$ as given by Eq.~(\ref{Eq:By_along_y}). Analogous to Eq.~(\ref{K_def}), the effective undulator parameter can be defined as
\begin{equation}
K_\textrm{eff} (y) \equiv \frac{e \lambda_w}{2 \pi m_e c} B_y (y, 0) = \frac{e \lambda_w}{2 \pi m_e c}  B_0 \cosh (k_w y) = K \cosh (k_w y).
\end{equation}
A plot of $K_\textrm{eff}$ versus $y$ is shown in Fig.~\ref{Keff_vs_y}(a). Note that $K_\textrm{eff} (y) > K$ for all $y \neq 0$, meaning that any trajectory error in the $y$-direction effectively \textit{increases} the undulator parameter from the desired value $K$. 

The difference between the effective undulator parameter $K_{\text{eff}}$ and the desired value $K$ can be expressed as
\begin{equation}
\label{Eq:Keff_minus_K}
\delta K = K_{\text{eff}} - K = K [\cosh(k_w y) - 1].
\end{equation}
This difference of $\delta K $ in an undulator module, in turn, leads to a phase mismatch in the drift section thereafter. In the drift section, there is a phase advance due to the speed difference between the electron beam and the radiation emitted in the preceding undulator module. For a drift length $L_D$ after an undulator module with parameter $K$, this phase advance is~\cite{PhaseShifters}
\begin{equation}
\label{Eq:phi}
\phi = k_w L_D \left( 1 + \frac{K^2}{2} \right)^{-1}.
\end{equation}
The phase shifter in the drift section is configured to perform automatic phase matching for the $\phi$ associated with the input value $K$. Thus, the difference of $\delta K $ causes a phase shift error of
\begin{equation}
\label{Eq:deltaphi}
\delta \phi = - k_w L_D K \left( 1 + \frac{K^2}{2} \right)^{-2} \delta K = - k_w L_D K^2 \left( 1 + \frac{K^2}{2} \right)^{-2} [\cosh(k_w y) - 1].
\end{equation}
The absolute phase error $|\delta \phi|$ is shown in Fig.~\ref{Keff_vs_y}(b) as a function of $y$. In this discussion, the additional phase advance due to the fringe fields at the two ends of an undulator module is not considered.


\end{document}